\documentclass[aps,preprint]{revtex4}%
\usepackage{amsfonts}
\usepackage{amsmath}
\usepackage{amssymb}
\usepackage{graphicx}%
\begin{document}
\preprint{CTP-SCU/2016007}
\title{Quantum Tunneling In Deformed Quantum Mechanics with Minimal Length}
\author{Xiaobo Guo$^{a}$}
\email{guoxiaobo@swust.edu.cn}
\author{Bochen Lv$^{b}$}
\email{bochennn@yahoo.com}
\author{Jun Tao$^{b}$}
\email{taojun@scu.edu.cn}
\author{Peng Wang$^{b}$}
\email{pengw@scu.edu.cn}
\affiliation{$^{a}$School of Science, Southwest University of Science and Technology,
Mianyang, 621010, PR China}
\affiliation{$^{b}$Center for Theoretical Physics, College of Physical Science and
Technology, Sichuan University, Chengdu, 610064, PR China}

\begin{abstract}
{}In the deformed quantum mechanics with a minimal length, one WKB connection
formula through a turning point is derived. We then use it to calculate
tunnelling rates through potential barriers under the WKB approximation.
Finally, the minimal length effects on two examples of quantum tunneling in
nuclear and atomic physics are discussed.

\end{abstract}
\keywords{}\maketitle
\tableofcontents

\section{Introduction}

Various theories of quantum gravity, such as string theory, loop quantum
gravity and quantum geometry, predict the existence of a minimal length
\cite{IN-Townsend:1977xw,IN-Amati:1988tn,IN-Konishi:1989wk}. For a review of a
minimal length in quantum gravity, see \cite{IN-Garay:1994en}. Some
realizations of the minimal length from various scenarios have been proposed.
Specifically, one of the most popular models is the Generalized Uncertainty
Principle (GUP) \cite{IN-Maggiore:1993kv,IN-Kempf:1994su}, derived from the
modified fundamental commutation relation:
\begin{equation}
\lbrack X,P]=i\hbar(1+\beta P^{2}), \label{1dGUP}%
\end{equation}
where $\beta=\beta_{0}\ell_{p}^{2}/\hbar^{2}=\beta_{0}/c^{2}m_{pl}^{2}$,
$m_{pl}$ is the Planck mass, $\ell_{p}$ is the Planck length, and $\beta_{0}$
is a dimensionless parameter. With this modified commutation relation, one can
easily find
\begin{equation}
\Delta X\Delta P\geq\frac{\hbar}{2}[1+\beta(\Delta P)^{2}], \label{2dGUP}%
\end{equation}
which leads to the minimal measurable length:
\begin{equation}
\Delta X\geq\Delta_{\text{min}}=\hbar\sqrt{\beta}=\sqrt{\beta_{0}}\ell_{p}.
\label{GUPshixian-2}%
\end{equation}
The GUP has been extensively studied recently, see for example
\cite{IN-Chang:2001bm,IN-Brau:1999uv,IN-Das:2008kaa,IN-Hossenfelder:2003jz,IN-Ali:2009zq,IN-Li:2002xb,IN-Brau:2006ca,IN-Wang:2015bwa}%
. For a review of the GUP, see \cite{IN-Hossenfelder:2012jw}.

To study 1D quantum mechanics with the deformed commutators $(\ref{1dGUP})$,
one can exploit the following representation for $X$ and $P$:%
\begin{equation}
X=X_{0}\text{, }P=P_{0}\left(  1+\frac{\beta}{3}P_{0}^{2}\right)  ,
\end{equation}
where $[X_{0},P_{0}]=i\hbar$. It can easily show that such representation
fulfills the relation $(\ref{1dGUP})$ to $\mathcal{O}\left(  \beta\right)  $.
Furthermore, we can adopt the position representation:%
\begin{equation}
X_{0}=x\text{, }P_{0}=\frac{\hbar}{i}\frac{\partial}{\partial x}\text{.}%
\end{equation}
Therefore for a quantum system described by%
\begin{equation}
H=\frac{P^{2}}{2m}+V\left(  X\right)  ,
\end{equation}
the deformed stationary Schrodinger equation in the position representation is%
\begin{equation}
\frac{d^{2}\psi\left(  x\right)  }{dx^{2}}-\ell_{\beta}^{2}\frac{d^{4}%
\psi\left(  x\right)  }{dx^{4}}+\frac{2m\left[  E-V\left(  x\right)  \right]
}{\hbar^{2}}\psi\left(  x\right)  =0. \label{eq: Sch-eqn}%
\end{equation}
where $\ell_{\beta}^{2}\equiv\frac{2}{3}\hbar^{2}\beta$ and terms of order
$\beta^{2}$ are neglected.

If eqn. $\left(  \ref{eq: Sch-eqn}\right)  $ with $\beta=0$ can be solved
exactly, one could use the perturbation method to solve eqn. $\left(
\ref{eq: Sch-eqn}\right)  $ by treating the term with $\ell_{\beta}^{2}$ as a
small correction. However for the general $V\left(  x\right)  $, one might
need other methods to solve eqn. $\left(  \ref{eq: Sch-eqn}\right)  $. In
fact, the WKB approximation in deformed space have been considered
\cite{IN-Fityo:2005lwa}. In \cite{IN-Fityo:2005lwa}, the authors considered
the deformed commutation relation%
\begin{equation}
\left[  X,P\right]  =i\hbar f\left(  P\right)  , \label{eq:generalGUP}%
\end{equation}
where $f\left(  P\right)  $ is some function. For $f\left(  P\right)  $, one
could solve the differential equation%
\begin{equation}
\frac{dP\left(  p\right)  }{dp}=f\left(  P\right)
\end{equation}
for $P\left(  p\right)  $, and $p\left(  P\right)  $ denotes the inverse
function of $P\left(  p\right)  $. It is interesting to note that there might
be more than one inverse function for $P\left(  p\right)  $. However, one
usually finds that there is only one inverse function $p\left(  P\right)  $
which vanishes at $P=0$. The rest ones are called "runaway" solutions, which
are not physical and should be discarded \cite{IN-Simon:1990ic,IN-Mu:2015qna}.
Then, they used the WKB approximation to show that the solution to the
deformed Schrodinger equation:%
\begin{equation}
P^{2}\left(  \frac{\hbar}{i}\frac{d}{dx}\right)  \psi\left(  x\right)
+2m\left[  V\left(  x\right)  -E\right]  \psi\left(  x\right)  =0,
\label{eq:deformed-Sch-general}%
\end{equation}
was%
\begin{equation}
\psi\left(  x\right)  =\frac{1}{\sqrt{\left\vert P\left(  x\right)  f\left(
P\left(  x\right)  \right)  \right\vert }}\left(  C_{1}\exp\left[  \frac
{i}{\hbar}\int^{x}p\left(  P\left(  x\right)  \right)  dx\right]  +C_{2}%
\exp\left[  -\frac{i}{\hbar}\int^{x}p\left(  P\left(  x\right)  \right)
dx\right]  \right)  , \label{eq:WKB-sol}%
\end{equation}
where $P\left(  x\right)  =\sqrt{2m\left[  E-V\left(  x\right)  \right]  }$.
Moreover, it also showed that the condition
\begin{equation}
\left\vert P^{2}\left(  x\right)  \right\vert \gg\hbar\left\vert \frac{d}%
{dx}P\left(  x\right)  f\left(  P\left(  x\right)  \right)  \right\vert ,
\label{eq:Condition}%
\end{equation}
had to be satisfied to make the WKB approximation valid. However, the
condition $\left(  \ref{eq:Condition}\right)  $ fails near a turning point
where $P\left(  x\right)  =0$.

For the case with $f\left(  P\right)  =1+\beta P^{2}$, we derived one WKB
connection formula through turning points and Bohr-Sommerfeld quantization
rule in \cite{IN-Tao:2012fp}. In this paper, we continue to consider other WKB
connection formulas and calculate tunnelling rates through potential barriers.
The remainder of our paper is organized as follows. In section \ref{Sec:TTPB},
we derive one WKB connection formula and use it to find the formula for the
tunnelling rate through a potential barrier. Then two examples of quantum
tunneling in nuclear and atomic physics are discussed in section
\ref{Sec:App}. Section \ref{Sec:Con} is devoted to our conclusions.

\section{Tunneling Through Potential Barriers}

\label{Sec:TTPB}

\begin{figure}[tb]
\begin{centering}
\includegraphics[scale=0.8]{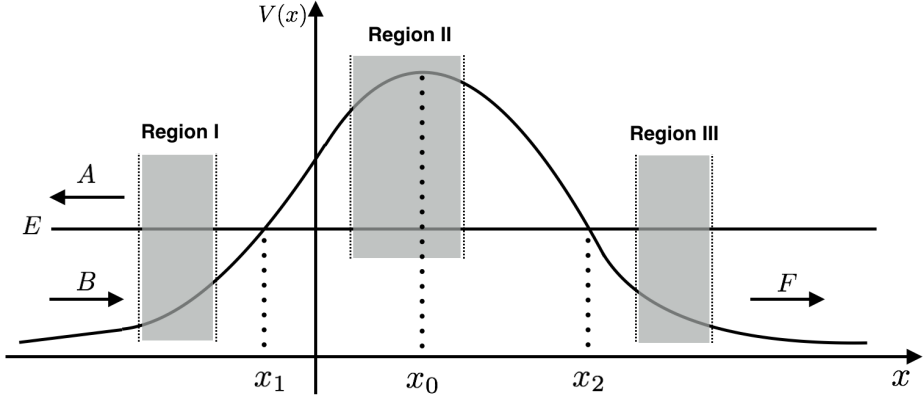}
\par\end{centering}
\caption{Scattering from a potential barrier.}%
\label{fig:potential}%
\end{figure}

We now consider WKB description of tunneling through a potential barrier
$V\left(  x\right)  $, which vanishes as $x\rightarrow\pm\infty$ and rises
monotonically to its maximum $V_{\max}$ at $x=x_{0}$ as $x$ approaches $x_{0}$
from either the left or the right side of $x_{0}$. In FIG.
$\ref{fig:potential}$, we plot the potential $V\left(  x\right)  $. For a
particle of energy $E$, there are two turning points $x=x_{1}\ $and $x=x_{2}$,
$x_{1}<x_{2}$, at which $V\left(  x\right)  =E$. There are two classical
allowed regions, Region I with $x<x_{1}$ and Region III with $x>x_{2}$. To
describe tunneling, we need to choose appropriate boundary conditions in the
classical allowed regions. We postulate an incident right-moving wave in
Region I, where the WKB approximation solution to eqn. $\left(
\ref{eq:deformed-Sch-general}\right)  $ includes a wave incident the barrier
and a reflected wave:%
\begin{equation}
\psi\left(  x\right)  =\frac{A\exp\left[  -\frac{i}{\hbar}\int_{x}^{x_{1}%
}p\left(  P\left(  x\right)  \right)  dx\right]  }{\sqrt{\left\vert P\left(
x\right)  f\left(  P\left(  x\right)  \right)  \right\vert }}+\frac
{B\exp\left[  \frac{i}{\hbar}\int_{x}^{x_{1}}p\left(  P\left(  x\right)
\right)  dx\right]  }{\sqrt{\left\vert P\left(  x\right)  f\left(  P\left(
x\right)  \right)  \right\vert }}.
\end{equation}
In Region III, there is only a transmitted wave:%
\begin{equation}
\psi\left(  x\right)  =\frac{F\exp\left[  \frac{i}{\hbar}\int_{x_{2}}%
^{x}p\left(  P\left(  x\right)  \right)  dx\right]  }{\sqrt{\left\vert
P\left(  x\right)  f\left(  P\left(  x\right)  \right)  \right\vert }}.
\label{eq:Trans-WKB}%
\end{equation}
In the classically forbidden Region II, there are exponentially growing and
decaying solutions:%
\begin{align}
\psi\left(  x\right)   &  =\frac{C\exp\left[  -\frac{1}{\hbar}\int_{x_{1}}%
^{x}\left\vert p\left(  P\left(  x\right)  \right)  \right\vert dx\right]
}{\sqrt{\left\vert P\left(  x\right)  f\left(  P\left(  x\right)  \right)
\right\vert }}+\frac{D\exp\left[  \frac{1}{\hbar}\int_{x_{1}}^{x}\left\vert
p\left(  P\left(  x\right)  \right)  \right\vert dx\right]  }{\sqrt{\left\vert
P\left(  x\right)  f\left(  P\left(  x\right)  \right)  \right\vert }%
}\nonumber\\
&  =\frac{C^{\prime}\exp\left[  \frac{1}{\hbar}\int_{x}^{x_{2}}\left\vert
p\left(  P\left(  x\right)  \right)  \right\vert dx\right]  }{\sqrt{\left\vert
P\left(  x\right)  f\left(  P\left(  x\right)  \right)  \right\vert }}%
+\frac{D^{\prime}\exp\left[  -\frac{1}{\hbar}\int_{x}^{x_{2}}\left\vert
p\left(  P\left(  x\right)  \right)  \right\vert dx\right]  }{\sqrt{\left\vert
P\left(  x\right)  f\left(  P\left(  x\right)  \right)  \right\vert }},
\label{eq:Forb-WKB}%
\end{align}
where
\begin{align}
C^{\prime}  &  =Ce^{-\eta},D^{\prime}=De^{\eta},\nonumber\\
\eta &  \equiv\frac{1}{\hbar}\int_{x_{1}}^{x_{2}}\left\vert p\left(  P\left(
x\right)  \right)  \right\vert dx. \label{eq:CD-CDprime}%
\end{align}
To calculate the tunneling rate, we need to use connection formulas to relate
$F$, $C/C^{\prime}$, and $D/D^{\prime}$ to $A$. In \cite{IN-Tao:2012fp}, we
derived one WKB connection formula around $x=x_{1}$ in the case with $f\left(
P\right)  =1+\beta P^{2}$. If $D=0$, we found that%
\begin{align}
&  \frac{C}{\sqrt{\left\vert P\left(  x\right)  f\left(  P\left(  x\right)
\right)  \right\vert }}\exp\left(  -\frac{1}{\hbar}\int_{x_{1}}^{x}\left\vert
p\left(  P\left(  x\right)  \right)  \right\vert dx\right) \nonumber\\
&  \rightarrow\frac{2C}{\sqrt{\left\vert P\left(  x\right)  f\left(  P\left(
x\right)  \right)  \right\vert }}\sin\left(  \frac{1}{\hbar}\int_{x_{1}}%
^{x}p\left(  P\left(  x\right)  \right)  dx+\frac{\pi}{4}\right)  \text{, to
}\mathcal{O}\left(  \beta^{2}\right)  , \label{eq:connectionA}%
\end{align}
which gives
\begin{equation}
A=Ce^{-\frac{i\pi}{4}}. \label{eq:A-C}%
\end{equation}
In what follows, we will derive a WKB connection formula around $x=x_{2}$ to
relate $C^{\prime}$ and $D^{\prime}$ to $F$ and then calculate the tunneling
rate through the potential barrier.

To match WKB solutions, we need to solve the deformed Schrodinger equation
$\left(  \ref{eq: Sch-eqn}\right)  $ in the vicinity of the turning point
$x=x_{2}$. A linear approximation to the potential $V\left(  x\right)  $\ near
the turning point $x=x_{2}$ is
\begin{equation}
V\left(  x\right)  \approx V\left(  x_{2}\right)  -\left\vert V^{\prime
}\left(  x_{2}\right)  \right\vert \left(  x-x_{2}\right)  , \label{linearV}%
\end{equation}
where $V\left(  x_{2}\right)  =E$. To simplify eqn. $\left(  \ref{eq: Sch-eqn}%
\right)  $, a new dimensionless variable $\rho$ could be introduced:%
\begin{equation}
\rho=\left(  x_{2}-x\right)  \left(  2m\left\vert V^{\prime}\left(
x_{2}\right)  \right\vert /\hbar^{2}\right)  ^{\frac{1}{3}}. \label{eq:rho-x}%
\end{equation}
Thus, eqn. $\left(  \ref{eq: Sch-eqn}\right)  $ becomes%
\begin{equation}
-\alpha^{2}\frac{d^{4}\psi\left(  \rho\right)  }{d\rho^{4}}+\frac{d^{2}%
\psi\left(  \rho\right)  }{d\rho^{2}}-\rho\psi\left(  \rho\right)  =0,
\label{eq:Sch-eqn-rho}%
\end{equation}
where $\alpha^{2}=\left(  2m\left\vert V^{\prime}\left(  x_{2}\right)
\right\vert /\hbar^{2}\right)  ^{\frac{2}{3}}\ell_{\beta}^{2}$. The
differential equation $\left(  \ref{eq: Sch-eqn}\right)  $ can be solved by
Laplace's method \cite{IN-Tao:2012fp}. Integral representations of the
solutions are%
\begin{equation}
I\left(  \rho\right)  =\int_{C}\exp\left(  \rho t+\frac{\alpha^{2}t^{5}}%
{5}-\frac{t^{3}}{3}\right)  dt, \label{eq:solution}%
\end{equation}
where the contour $C$ is chosen so that the integrand vanishes at endpoints of
$C$. Specifically, define five sectors:
\begin{equation}
\Theta_{n}\equiv\left[  \frac{2n\pi+\frac{\pi}{2}}{5},\frac{2n\pi+\frac{3\pi
}{2}}{5}\right]  ,\text{ }n=0,1,2,3,4.
\end{equation}
The contour $C$ must originate at one of them and terminate at another.

The asymptotic expressions of $I\left(  \rho\right)  $ for large values of
$\rho$ can be obtained by evaluating the integral $\left(  \ref{eq:solution}%
\right)  $ using the method of steepest descent. To do this, we make the
change of variables $t=\left\vert \rho\right\vert ^{\frac{1}{2}}s$ and find%
\begin{equation}
I\left(  \rho\right)  =\left\vert \rho\right\vert ^{\frac{1}{2}}\int_{C}%
\exp\left[  \left\vert \rho\right\vert ^{\frac{3}{2}}f_{\pm}\left(  s\right)
\right]  ds, \label{eq:int-rho}%
\end{equation}
where $f_{\pm}\left(  s\right)  $ $\equiv\pm s-\frac{s^{3}}{3}+\frac{as^{5}%
}{5}$ with $+$ for $\rho>0$ and $-$ for $\rho<0$, and $a\equiv\alpha
^{2}\left\vert \rho\right\vert \ll1$ in the physical region
\cite{IN-Tao:2012fp}. We will show below that there exists a steepest descent
contour ranging from $s=\infty\exp\left(  \frac{7\pi i}{5}\right)  $ to
$s=\infty\exp\left(  \frac{9\pi i}{5}\right)  $, which is the red contour in
FIG. $\ref{fig:negative}$. Such contour could let us match the asymptotic
expression of $I\left(  \rho\right)  $ at large negative value of $\rho$ with
the WKB solution $\left(  \ref{eq:Trans-WKB}\right)  $ in Region III. Note
that $\infty\exp\left(  \frac{7\pi i}{5}\right)  \in\Theta_{3}$ and
$\infty\exp\left(  \frac{7\pi i}{5}\right)  \in\Theta_{4}$.

The method of steepest descent is very powerful to calculate integrals of the
form%
\begin{equation}
I\left(  x\right)  =\int_{C}g\left(  z\right)  e^{xf\left(  z\right)  }dz,
\end{equation}
where $C$ is a contour in the complex plane. We are usually interested in the
behavior of $I\left(  x\right)  $ as $x\rightarrow\infty$. The key step of the
method of steepest descent is applying Cauchy's theorem to deform the contours
$C$ to the contours coinciding with steepest descent paths. Around a saddle
point $z_{0}$ where $f^{\prime}\left(  z_{0}\right)  =0$, there are two
cosntant-phase (steepest) contours, on which $\operatorname{Im}f\left(
z\right)  $ is constant, passing through $z_{0}$ if $f^{\prime\prime}\left(
z_{0}\right)  \neq0$. One of them is a steepest descent contour, along which
$\operatorname{Re}f\left(  z\right)  $ increases as we go towards $z_{0}$. The
other is a steepest ascent contour, along which $\operatorname{Re}f\left(
z\right)  $ decreases as we go towards $z_{0}$. If $I\left(  x\right)  $ is
integrated along the steepest descent contour, the asymptotic behavior of
$I\left(  x\right)  $ is dominated by the contribution from the saddle point
$z_{0}$.

\begin{figure}[tb]
\begin{centering}
\includegraphics[scale=0.8]{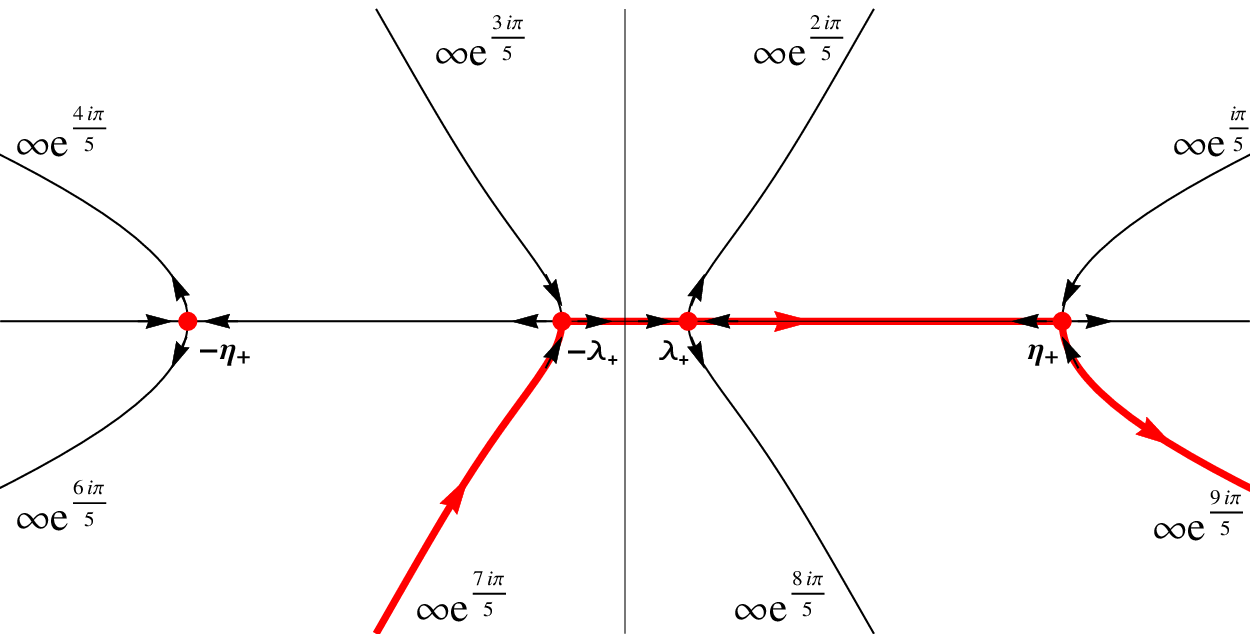}
\par\end{centering}
\caption{Saddle points and cosntant-phase (steepest) contours of $f_{+}\left(
s\right)  $. The red contour is a steepest descent contour, along which
$I\left(  \rho\right)  $ is integrated. The black arrows on the cosntant-phase
contours around saddle points denote the directions in which values of
$\operatorname{Re}f_{+}\left(  s\right)  $ increase.}%
\label{fig:positive}%
\end{figure}

\begin{figure}[tb]
\begin{centering}
\includegraphics[scale=0.8]{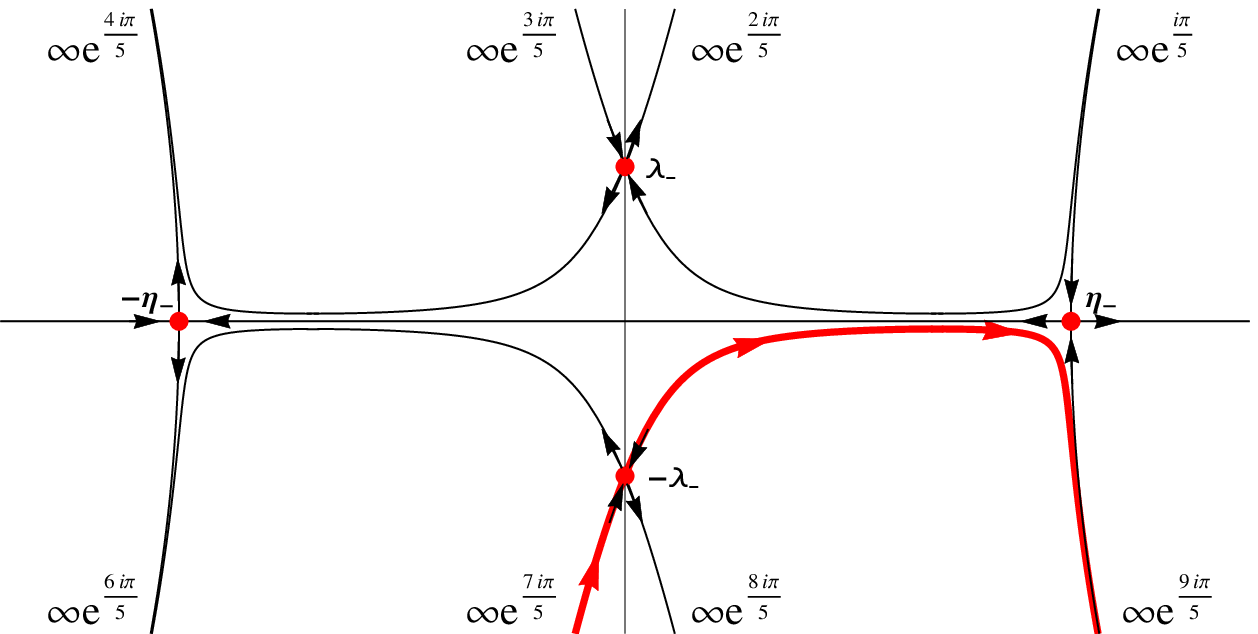}
\par\end{centering}
\caption{Saddle points and cosntant-phase (steepest) contours of $f_{-}\left(
s\right)  $. The red contour is a steepest descent contour, along which
$I\left(  \rho\right)  $ is integrated. The black arrows on the cosntant-phase
contours around saddle points denote the directions in which values of
$\operatorname{Re}f_{-}\left(  s\right)  $ increase.}%
\label{fig:negative}%
\end{figure}

In FIGs. $\ref{fig:positive}$ and $\ref{fig:negative}$, we plot saddle points
(red points in figures) of $f_{+}\left(  s\right)  $ and $f_{-}\left(
s\right)  $, respectively, and cosntant-phase contours passing through them.
Specifically, saddle points of $f_{+}\left(  s\right)  $ are%
\begin{equation}
s=\pm\lambda_{+}\equiv\pm\frac{\sqrt{1-\sqrt{1-4a}}}{\sqrt{2a}}\text{ and
}s=\pm\eta_{+}\equiv\pm\frac{\sqrt{1+\sqrt{1-4a}}}{\sqrt{2a}},
\end{equation}
and these of $f_{-}\left(  s\right)  $ are
\begin{equation}
s=\pm\lambda_{-}\equiv\pm\frac{\sqrt{1-\sqrt{1+4a}}}{\sqrt{2a}}\text{ and
}s=\pm\eta_{-}\equiv\pm\frac{\sqrt{1+\sqrt{1+4a}}}{\sqrt{2a}}%
.\label{saddlepointminus}%
\end{equation}
The red contours in FIGs. $\ref{fig:positive}$ and $\ref{fig:negative}$ are
the steepest descent contours connecting $s=\infty\exp\left(  \frac{7\pi i}%
{5}\right)  $ to $s=\infty\exp\left(  \frac{9\pi i}{5}\right)  $, along which
the integral $\left(  \ref{eq:int-rho}\right)  $ is integrated. Note that red
arrows on them denote the steepest contours' directions. On the other hand,
the black arrows on the cosntant-phase contours around saddle points denote
the directions in which values of $\operatorname{Re}f_{\pm}\left(  s\right)  $
increase. Following the black arrows on the red contour in FIG.
$\ref{fig:positive}$, we find that $\operatorname{Re}f_{+}\left(  -\lambda
_{+}\right)  $ and $\operatorname{Re}f_{+}\left(  \eta_{+}\right)  $ are
smaller than $\operatorname{Re}f_{+}\left(  \lambda_{+}\right)  $. Thus for
the case with $\rho>0$, the asymptotic expression of $I\left(  \rho\right)  $
is dominated by the contribution from the saddle $s=\lambda_{+}$. The method
of steepest descent gives
\begin{align}
I\left(  1\ll\rho\ll\alpha^{-2}\right)   &  \sim\frac{\sqrt{\pi}\exp\left[
\rho^{\frac{3}{2}}f_{+}\left(  \lambda_{+}\right)  \right]  }{\rho^{\frac
{1}{4}}}\sqrt{\frac{2}{\left\vert f_{+}^{\prime\prime}\left(  \lambda
_{+}\right)  \right\vert }}\nonumber\\
&  \sim\frac{\sqrt{\pi}\left(  1+\frac{3}{4}a\right)  }{\rho^{\frac{1}{4}}%
}\exp\left[  \frac{2\rho^{\frac{3}{2}}}{3}\left(  1+\frac{3a}{10}\right)
\right]  ,\label{eq:I-largeP-rho}%
\end{align}
where $a\ll1$ is used, and terms of $\mathcal{O}\left(  a^{2}\right)  $ are
neglected in the second line. For the case with $\rho>0$, FIG.
$\ref{fig:negative}$ shows that the asymptotic expression of $I\left(
\rho\right)  $ is dominated by the contribution from the saddle $s=-\lambda
_{-}$, and hence we find%
\begin{align}
I\left(  -1\gg\rho\gg-\alpha^{-2}\right)   &  \sim\frac{\sqrt{\pi}\exp\left(
\frac{\pi}{4}i\right)  \exp\left[  \left\vert \rho\right\vert ^{\frac{3}{2}%
}f_{-}\left(  -\lambda_{-}\right)  \right]  }{\left\vert \rho\right\vert
^{\frac{1}{4}}}\sqrt{\frac{2}{\left\vert f_{-}^{\prime\prime}\left(
-\lambda_{+}\right)  \right\vert }}\nonumber\\
&  \sim\frac{\sqrt{\pi}\exp\left(  \frac{\pi}{4}i\right)  \left(  1-\frac
{3}{4}a\right)  }{\left\vert \rho\right\vert ^{\frac{1}{4}}}\exp\left[
\frac{2i\left\vert \rho\right\vert ^{\frac{3}{2}}}{3}\left(  1-\frac{3a}%
{10}\right)  \right]  ,\label{eq:I-largeN-rho}%
\end{align}
where terms of $\mathcal{O}\left(  a^{2}\right)  $ are neglected in the second line.

Around the turning point $x=x_{2}$, $\left\vert x-x_{2}\right\vert \ll1$ and
$P\left(  x\right)  \sim\sqrt{2m\left\vert V^{\prime}\left(  x_{2}\right)
\right\vert }\sqrt{x-x_{2}}$. In this region, we find that WKB solutions
$\left(  \ref{eq:Trans-WKB}\right)  $ and $\left(  \ref{eq:Forb-WKB}\right)  $
become%
\begin{align}
\psi\left(  x\right)   &  \sim\frac{F}{\left(  2m\left\vert V^{\prime}\left(
x_{2}\right)  \right\vert /\hbar^{2}\right)  ^{\frac{1}{3}}\hbar}\frac
{1-\frac{3a}{4}}{\left\vert \rho\right\vert ^{\frac{1}{4}}}\exp\left[
\frac{2i\left\vert \rho\right\vert ^{\frac{3}{2}}}{3}\left(  1-\frac{3a}%
{10}\right)  \right]  ,\nonumber\\
\psi\left(  x\right)   &  \sim\frac{1}{\left(  2m\left\vert V^{\prime}\left(
x_{2}\right)  \right\vert /\hbar^{2}\right)  ^{\frac{1}{3}}\hbar}\frac
{1+\frac{3}{4}a}{\rho^{\frac{1}{4}}}\left\{  C^{\prime}\exp\left[  \frac
{2\rho^{\frac{3}{2}}}{3}\left(  1+\frac{3a}{10}\right)  \right]  +D^{\prime
}\exp\left[  -\frac{2\rho^{\frac{3}{2}}}{3}\left(  1+\frac{3a}{10}\right)
\right]  \right\}  , \label{eq:WKB-Aprro}%
\end{align}
where we use $p\left(  P\right)  \approx P-\frac{\beta}{3}P^{3}$ and terms of
$\mathcal{O}\left(  a^{2}\right)  $ are neglected, and we express $x$ in terms
of $\rho$ using eqn. $\left(  \ref{eq:rho-x}\right)  $. In the overlap regions
where $\left\vert \rho\right\vert \gg1$ and $\left\vert x-x_{2}\right\vert
\ll1$, matching WKB solutions $\left(  \ref{eq:WKB-Aprro}\right)  $ with the
$I\left(  \rho\right)  $'s asymptotic expressions $\left(
\ref{eq:I-largeP-rho}\right)  $ and $\left(  \ref{eq:I-largeN-rho}\right)  $
gives%
\begin{equation}
C^{\prime}=F\exp\left(  -\frac{\pi}{4}i\right)  \text{ and }D^{\prime
}=0\text{,}%
\end{equation}
which by eqns. $\left(  \ref{eq:CD-CDprime}\right)  $ lead to $C=F\exp\left(
-\frac{\pi}{4}i\right)  e^{\eta}$ and $D=0$. Since $D=0$, eqn. $\left(
\ref{eq:A-C}\right)  $ gives%
\begin{equation}
F=iAe^{-\eta},
\end{equation}
and the transmission probability is%
\begin{equation}
T=\frac{\left\vert F\right\vert ^{2}}{\left\vert A\right\vert ^{2}}\sim
e^{-2\eta}. \label{eq:tunneling}%
\end{equation}

\section{Examples}

\label{Sec:App}

The dimensionless number $\beta_{0}=c^{2}m_{pl}^{2}\beta=\hbar^{2}\beta
/\ell_{p}^{2}$ plays an important role when implications and applications of
non-zero minimal length are discussed. Normally, if the minimal length is
assumed to be order of the Planck length $\ell_{p}$, one has $\beta_{0}\sim1$.
In \cite{IN-Das:2008kaa}, based on the precision measurement of STM current,
an upper bound of $\beta_{0}$ was given by $\beta_{0}<10^{21}$. In the
following, we use eqn. $\left(  \ref{eq:tunneling}\right)  $ to study effects
of GUP on $\alpha$ decay and cold electrons emission from metal via strong
external electric field.

\subsection{$\alpha$ Decay}

\begin{figure}[tb]
\begin{centering}
\includegraphics[scale=0.6]{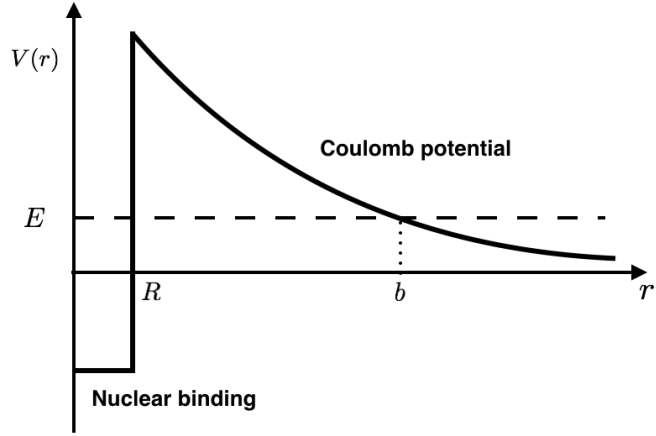}
\par\end{centering}
\caption{The potential energy of an $\alpha$-particle in a radioactive
nucleus.}%
\label{fig:alpga}%
\end{figure}

The decay of a nucleus into an $\alpha$-particle (charge $2e$) and a daughter
nucleus (charge $Ze$) can be described as the tunneling of an $\alpha
$-particle through a barrier caused by the Coulomb potential between the
daughter and the $\alpha$-particle (FIG. $\ref{fig:alpga}$)
\cite{APP-Gamow:1928zz}. For an $\alpha$-particle of energy $E$ in the
potential in FIG. $\ref{fig:alpga}$, there are two turning points, the nuclear
radius $R$ and the outer turning point $b$, which is determined by
\begin{equation}
E=\frac{Ze^{2}}{2\pi\varepsilon_{0}r}\Rightarrow b=\frac{Ze^{2}}%
{2\pi\varepsilon_{0}E}.
\end{equation}
The exponent $\eta$ in eqn. $\left(  \ref{eq:tunneling}\right)  $ is%
\begin{align}
\eta &  =\frac{1}{\hbar}\int_{R}^{b}\left\vert p\left(  \sqrt{2mE}%
\sqrt{1-\frac{Z_{1}Z_{2}e^{2}}{4\pi\varepsilon_{0}Er}}\right)  \right\vert
dx\nonumber\\
&  \approx\frac{\sqrt{2mE}}{\hbar}\left\{  b\arccos\left(  \sqrt{\frac{R}{b}%
}\right)  -\sqrt{R\left(  b-R\right)  }\right. \nonumber\\
&  \left.  +\frac{2mE\beta}{3}\left[  3b\operatorname{arcsec}\sqrt{\frac{b}%
{R}}-\sqrt{R\left(  b-R\right)  }-2b\sqrt{\frac{b-R}{R}}\right]  \right\}  ,
\end{align}
where $m$ is the mass of the $\alpha$-particle. At low energies (relative to
the height of the Coulomb barrier at $r=R$), we have $b\gg R$ and then%
\begin{equation}
\eta\approx\frac{\sqrt{2m}e^{2}}{4\varepsilon_{0}\hbar}\left[  1-\frac
{8mE\beta}{3\pi}\sqrt{\frac{b}{R}}\right]  \frac{Z}{\sqrt{E}}.
\end{equation}
The probability of emission of an $\alpha$-particle is proportional to
$e^{-2\eta}$ and hence the lifetime of the parent nucleus is about%
\begin{equation}
\tau\sim e^{2\eta}.
\end{equation}
The density of nuclear matter is relatively constant, so $R^{3}$ is
proportional to the number of nucleons $A$. Empirically, we have%
\begin{equation}
R\sim A^{\frac{1}{3}}\text{fm.}%
\end{equation}
Therefore, we find%
\begin{equation}
\ln\tau^{-1}\approx\text{const}-\frac{\sqrt{2m}e^{2}}{2\varepsilon_{0}\hbar
}\left[  1-\beta_{0}\left(  \frac{E}{\text{MeV}}\right)  ^{\frac{1}{2}}%
\sqrt{ZA^{-\frac{1}{3}}}\times10^{-40}\right]  \frac{Z}{\sqrt{E}}.
\label{eq:eta}%
\end{equation}
On the other hand, a large collection of data shows that a good fit to the
lifetime data obeys the Geiger--Nuttall law \cite{APP-Geiger}%
\begin{equation}
\ln\tau^{-1}=C_{1}-C_{2}\frac{Z}{\sqrt{E}},
\end{equation}
where $C_{1}$ and $C_{2}$ are constants. If the effects of GUP does not make
eqn. $\left(  \ref{eq:eta}\right)  $ differ too much from the Geiger--Nuttall
law, it will put an upper bound%
\begin{equation}
\beta_{0}<10^{40}.
\end{equation}
The GUP correction to the $\alpha$-decay has also been considered in
\cite{APP-Blado:2015ada}. We both find that the effects of the GUP would
increase the tunneling probability and hence decrease the lifetime $\tau$.

\subsection{Electron Emission from the Surface of Cold Metals}

\begin{figure}[tb]
\begin{centering}
\includegraphics[scale=0.6]{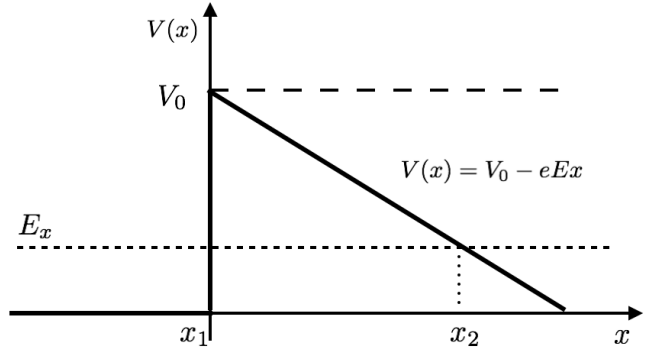}
\par\end{centering}
\caption{The potential energy inside and outside of a metallic surface when an
external electric field $E$ is added.}%
\label{fig:TP}%
\end{figure}

If a metal is placed in a very strong electric field, then there exists cold
emission of electrons from the surface of the metal. This emission of the
electrons can be explained via quantum tunneling. In \cite{APP-Fowler:1928bv},
the shape of a tunneling barrier was assumed to be the exact triangular
barrier, which has been known as the Fowler-Nordheim tunnelling. Note that
work must be done to remove an electron from the surface of a metal. In "free
electron gas" model, one could hence take the potential energy of the electron
inside the metal to be zero and for the outside to be $V\left(  x\right)
=V_{0}>0$. At the absolute zero temperature, if the Fermi energy of these
electrons $E_{F}$ is less than $V_{0}$, therefore after reaching the surface
of the metal, they are reflected back into the metal. Now if the external
electric field is applied toward the surface of the metal, the potential
energy becomes%
\begin{equation}
V\left(  x\right)  =\left\{
\begin{array}
[c]{c}%
V_{0}-eEx\text{ \ for }x>0\\
0\text{ \ \ \ \ \ \ \ \ for }x<0
\end{array}
\right.  ,
\end{equation}
where $E$ is the magnitude of the electric field. This potential is shown in
FIG. $\ref{fig:TP}$.

We now use eqn. $\left(  \ref{eq:tunneling}\right)  $ to calculate the GUP
modified transmission probability. For an electron of energy $E_{x}\leq
E_{F}<V_{0}$, there are two turning points:
\begin{equation}
x_{1}=0\text{ and }x_{2}=\frac{V_{0}-E_{x}}{eE}.
\end{equation}
The exponent $\eta$ in eqn. $\left(  \ref{eq:tunneling}\right)  $ is%
\begin{align}
\eta\left(  E_{x}\right)   &  \equiv\frac{1}{\hbar}\int_{x_{1}}^{x_{2}%
}\left\vert p\left(  \sqrt{2m\left[  E_{x}-V_{0}+eEx\right]  }\right)
\right\vert dx\nonumber\\
&  \approx\frac{2\sqrt{2m}\left(  V_{0}-E_{x}\right)  ^{\frac{3}{2}}}%
{3eE\hbar}\left[  1-\frac{2m\beta\left(  V_{0}-E_{x}\right)  }{5}\right]  ,
\end{align}
which gives the transmission probability $T\left(  E_{x}\right)  \approx
e^{-2\eta\left(  E_{x}\right)  }$.

Next we want to calculate the electric current density in this case. As a
consequence of the GUP, the number of quantum states should be changed to
\cite{APP-Wang:2010ct}
\begin{equation}
\frac{Vd^{3}p}{\left(  2\pi\hbar\right)  ^{3}\left(  1+\beta p^{2}\right)
^{3}},
\end{equation}
where $p^{2}=p_{i}p^{i}$. Therefore, the electric current density is given by%
\begin{equation}
J=e\int\frac{2dp_{x}dp_{y}dp_{z}}{\left(  2\pi\hbar\right)  ^{3}\left(
1+\beta p^{2}\right)  ^{3}}\frac{p_{x}}{m}T\left(  E_{x}\right)
\label{eq:current}%
\end{equation}
where $E_{x}=\frac{p_{x}^{2}}{2m}$. The range of $p_{x}$, $p_{y}$, are $p_{z}$
are inside the Fermi sphere:%
\begin{equation}
p_{x}^{2}+p_{y}^{2}+p_{z}^{2}\leq2mE_{F}.
\end{equation}
To calculate $J$, we use cylindrical coordinates:%
\begin{equation}
p_{y}=\rho\cos\phi\text{, }p_{z}=\rho\sin\phi\text{, and }\rho^{2}+p_{x}%
^{2}\leq2mE_{F},
\end{equation}
and have%
\begin{align}
J  &  =\frac{4\pi e}{\left(  2\pi\hbar\right)  ^{3}}\int_{0}^{\sqrt{2mE_{F}}%
}\frac{p_{x}}{m}T\left(  E_{x}\right)  dp_{x}\int_{0}^{\sqrt{2mE_{F}-p_{x}%
^{2}}}\frac{\rho d\rho}{\left[  1+\beta\left(  \rho^{2}+p_{x}^{2}\right)
\right]  ^{3}}\nonumber\\
&  \approx\frac{2\pi e}{\left(  2\pi\hbar\right)  ^{3}}\int_{0}^{\sqrt
{2mE_{F}}}\frac{p_{x}}{m}T\left(  E_{x}\right)  \left(  2mE_{F}-p_{x}%
^{2}\right)  \left[  1-\frac{3\beta}{2}\left(  2mE_{F}+p_{x}^{2}\right)
\right]  dp_{x}.
\end{align}
To simplify the result, we change $E_{x}$ to
\begin{equation}
\epsilon=E_{F}-E_{x}.
\end{equation}
Therefore, one has%
\begin{equation}
J\approx\frac{em}{2\pi^{2}\hbar^{3}}\int_{0}^{E_{F}}\epsilon T\left(
\epsilon\right)  \left[  1-3\beta m\left(  2E_{F}-\epsilon\right)  \right]
d\epsilon. \label{eq:J}%
\end{equation}
Since $T\left(  \epsilon\right)  $ decreases rapidly with increasing
$\epsilon$, therefore in $T\left(  \epsilon\right)  $ we can expand $\left(
V_{0}-E_{F}+\epsilon\right)  ^{\frac{3}{2}}$:%
\begin{equation}
\left(  V_{0}-E_{F}+\epsilon\right)  ^{\frac{3}{2}}\approx\left(  V_{0}%
-E_{F}\right)  ^{\frac{3}{2}}+\frac{3}{2}\epsilon\left(  V_{0}-E_{F}\right)
^{\frac{1}{2}}.
\end{equation}
We find%
\begin{equation}
J\approx\frac{em}{2\pi^{2}\hbar^{3}}\exp\left[  -\frac{2Q}{3}\right]
\frac{\left(  V_{0}-E_{F}\right)  ^{2}}{Q^{2}}\left[  1+\frac{4\beta m\left(
V_{0}-E_{F}\right)  Q}{15}\right]  , \label{eq:JJ}%
\end{equation}
where we extend the range of integration in eqn. $\left(  \ref{eq:J}\right)  $
to $\left(  0,\infty\right)  $, and
\begin{equation}
Q=\frac{2\sqrt{2m}\left(  V_{0}-E_{F}\right)  ^{\frac{3}{2}}}{\hbar eE}.
\end{equation}

In \cite{APP-Depas}, the Fowler-Nordheim tunneling in device grade ultra-thin
$3$-$6$ nm $n^{+}$poly-Si/SiO$_{2}$/$n$-Si structures has been analyzed.
Typically for this electron tunnelling, we have%
\begin{equation}
Q\sim10\text{, }V_{0}-E_{F}\sim1\text{eV, and }m\sim0.5\text{MeV.}%
\end{equation}
Therefore from eqn. $\left(  \ref{eq:JJ}\right)  $, the correction due to GUP
is given by
\begin{equation}
\frac{\delta J}{J}\sim10^{-50}\beta_{0}.
\end{equation}
The comparison of the calculated and experimental tunnel current was plotted
in FIG. 8 of \cite{APP-Depas}, which implies $\frac{\delta J}{J}%
\lesssim10^{-2}$. Then the upper bound on $\beta_{0}$ follows:%
\begin{equation}
\beta_{0}<10^{48}.
\end{equation}

\section{Conclusions}

\label{Sec:Con}

In this paper, we considered quantum tunneling in the deformed quantum
mechanics with a minimal length. We first found one WKB connection formula
through a turning point. Then the tunnelling rates through potential barriers
were derived using the WKB approximation. Finally, effects of the minimal
length on quantum tunneling were discussed in two examples in nuclear and
atomic physics, $\alpha$ decay and the Fowler-Nordheim tunnelling. Upper
bounds on $\beta_{0}$ were given in these two examples.

\begin{acknowledgments}
We are grateful to Houwen Wu and Zheng Sun for useful discussions. This work
is supported in part by NSFC (Grant No. 11005016, 11175039 and 11375121).
\end{acknowledgments}


\begin{thebibliography}{99}                                                                                               %


\bibitem {IN-Townsend:1977xw}P.~K.~Townsend, \textquotedblleft Small Scale
Structure of Space-Time as the Origin of the Gravitational
Constant,\textquotedblright\ Phys.\ Rev.\ D \textbf{15}, 2795 (1977). doi:10.1103/PhysRevD.15.2795

\bibitem {IN-Amati:1988tn}D.~Amati, M.~Ciafaloni and G.~Veneziano, ``Can
Space-Time Be Probed Below the String Size?,'' Phys.\ Lett.\ B \textbf{216},
41 (1989). doi:10.1016/0370-2693(89)91366-X

\bibitem {IN-Konishi:1989wk}K.~Konishi, G.~Paffuti and P.~Provero, ``Minimum
Physical Length and the Generalized Uncertainty Principle in String Theory,''
Phys.\ Lett.\ B \textbf{234}, 276 (1990). doi:10.1016/0370-2693(90)91927-4

\bibitem {IN-Garay:1994en}L.~J.~Garay, ``Quantum gravity and minimum length,''
Int.\ J.\ Mod.\ Phys.\ A \textbf{10}, 145 (1995) doi:10.1142/S0217751X95000085 [gr-qc/9403008].

\bibitem {IN-Maggiore:1993kv}M.~Maggiore, ``The Algebraic structure of the
generalized uncertainty principle,'' Phys.\ Lett.\ B \textbf{319}, 83 (1993)
doi:10.1016/0370-2693(93)90785-G [hep-th/9309034].

\bibitem {IN-Kempf:1994su}A.~Kempf, G.~Mangano and R.~B.~Mann, ``Hilbert space
representation of the minimal length uncertainty relation,'' Phys.\ Rev.\ D
\textbf{52}, 1108 (1995) doi:10.1103/PhysRevD.52.1108 [hep-th/9412167].

\bibitem {IN-Chang:2001bm}L.~N.~Chang, D.~Minic, N.~Okamura and T.~Takeuchi,
``The Effect of the minimal length uncertainty relation on the density of
states and the cosmological constant problem,'' Phys.\ Rev.\ D \textbf{65},
125028 (2002) doi:10.1103/PhysRevD.65.125028 [hep-th/0201017].

\bibitem {IN-Brau:1999uv}F.~Brau, ``Minimal length uncertainty relation and
hydrogen atom,'' J.\ Phys.\ A \textbf{32}, 7691 (1999)
doi:10.1088/0305-4470/32/44/308 [quant-ph/9905033].

\bibitem {IN-Das:2008kaa}S.~Das and E.~C.~Vagenas, ``Universality of Quantum
Gravity Corrections,'' Phys.\ Rev.\ Lett.\ \textbf{101}, 221301 (2008)
doi:10.1103/PhysRevLett.101.221301 [arXiv:0810.5333 [hep-th]].

\bibitem {IN-Hossenfelder:2003jz}S.~Hossenfelder, M.~Bleicher, S.~Hofmann,
J.~Ruppert, S.~Scherer and H.~Stoecker, ``Collider signatures in the Planck
regime,'' Phys.\ Lett.\ B \textbf{575}, 85 (2003)
doi:10.1016/j.physletb.2003.09.040 [hep-th/0305262].

\bibitem {IN-Ali:2009zq}A.~F.~Ali, S.~Das and E.~C.~Vagenas, ``Discreteness of
Space from the Generalized Uncertainty Principle,'' Phys.\ Lett.\ B
\textbf{678}, 497 (2009) doi:10.1016/j.physletb.2009.06.061 [arXiv:0906.5396 [hep-th]].

\bibitem {IN-Li:2002xb}X.~Li, ``Black hole entropy without brick walls,''
Phys.\ Lett.\ B \textbf{540}, 9 (2002) doi:10.1016/S0370-2693(02)02123-8 [gr-qc/0204029].

\bibitem {IN-Brau:2006ca}F.~Brau and F.~Buisseret, ``Minimal Length
Uncertainty Relation and gravitational quantum well,'' Phys.\ Rev.\ D
\textbf{74}, 036002 (2006) doi:10.1103/PhysRevD.74.036002 [hep-th/0605183].

\bibitem {IN-Wang:2015bwa}P.~Wang, H.~Yang and S.~Ying, \textquotedblleft
Minimal length effects on entanglement entropy of spherically symmetric black
holes in the brick wall model,\textquotedblright%
\ Class.\ Quant.\ Grav.\ \textbf{33}, no. 2, 025007 (2016)
doi:10.1088/0264-9381/33/2/025007 [arXiv:1502.00204 [gr-qc]].

\bibitem {IN-Hossenfelder:2012jw}S.~Hossenfelder, \textquotedblleft Minimal
Length Scale Scenarios for Quantum Gravity,\textquotedblright\ Living
Rev.\ Rel.\ \textbf{16}, 2 (2013) doi:10.12942/lrr-2013-2 [arXiv:1203.6191 [gr-qc]].

\bibitem {IN-Fityo:2005lwa}T.~V.~Fityo, I.~O.~Vakarchuk and V.~M.~Tkachuk,
``WKB approximation in deformed space with minimal length,'' J.\ Phys.\ A
\textbf{39}, no. 2, 379 (2006). doi:10.1088/0305-4470/39/2/0088

\bibitem {IN-Simon:1990ic}J.~Z.~Simon, ``Higher Derivative Lagrangians,
Nonlocality, Problems and Solutions,'' Phys.\ Rev.\ D \textbf{41}, 3720
(1990). doi:10.1103/PhysRevD.41.3720

\bibitem {IN-Mu:2015qna}B.~Mu, P.~Wang and H.~Yang, ``Thermodynamics and
Luminosities of Rainbow Black Holes,'' JCAP \textbf{1511}, no. 11, 045 (2015)
doi:10.1088/1475-7516/2015/11/045 [arXiv:1507.03768 [gr-qc]].

\bibitem {IN-Tao:2012fp}J.~Tao, P.~Wang and H.~Yang, ``Homogeneous Field and
WKB Approximation In Deformed Quantum Mechanics with Minimal Length,''
Adv.\ High Energy Phys.\ \textbf{2015}, 718359 (2015) doi:10.1155/2015/718359
[arXiv:1211.5650 [hep-th]].

\bibitem {APP-Gamow:1928zz}G.~Gamow, \textquotedblleft Zur Quantentheorie des
Atomkernes,\textquotedblright\ Z.\ Phys.\ \textbf{51}, 204 (1928). doi:10.1007/BF01343196

\bibitem {APP-Geiger}H. Geiger, J. M. Nuttall, \textquotedblleft The ranges of
the $\alpha$ particles from various radioactive substances and a relation
between range and period of transformation,\textquotedblright\ Philos. Mag.
22, 613 (1911).

\bibitem {APP-Blado:2015ada}G.~Blado, T.~Prescott, J.~Jennings, J.~Ceyanes and
R.~Sepulveda, ``Effects of the Generalized Uncertainty Principle on Quantum
Tunneling,'' Eur.\ J.\ Phys.\ \textbf{37}, 025401 (2016)
doi:10.1088/0143-0807/37/2/025401 [arXiv:1509.07359 [quant-ph]].

\bibitem {APP-Fowler:1928bv}R.~H.~Fowler and L.~Nordheim, ``Electron emission
in intense electric fields,'' Proc.\ Roy.\ Soc.\ Lond.\ A \textbf{119}, 173
(1928). doi:10.1098/rspa.1928.0091

\bibitem {APP-Wang:2010ct}P.~Wang, H.~Yang and X.~Zhang, \textquotedblleft
Quantum gravity effects on statistics and compact star
configurations,\textquotedblright\ JHEP \textbf{1008} (2010) 043
doi:10.1007/JHEP08(2010)043 [arXiv:1006.5362 [hep-th]].

\bibitem {APP-Depas}M. Depas, B. Vermeire, P. W. Mertens, R. L. Van Meirhaeghe
and M. M. Heyns, \textquotedblleft Determination of tunneling parameters in
ultra-thin oxide layer poly-Si/SiO2/Si structures,\textquotedblright%
\ Solid-State Electronics, vol. \textbf{38}, pp. 1465-1471, Aug. 1995.
\end{thebibliography}
\end{document}